\documentclass[aps,pra,showpacs,twocolumn,longbibliography]{revtex4-1}
\usepackage[utf8]{inputenc}
\usepackage{color}
\definecolor{c1}{rgb}{0.267004, 0.004874, 0.329415}
\definecolor{c2}{rgb}{0.192357, 0.403199, 0.555836}
\usepackage[colorlinks,pdfusetitle,urlcolor=c1,citecolor=c1,linkcolor=c2,bookmarksnumbered,plainpages=false]{hyperref}
\usepackage{amsfonts}
\usepackage{amssymb}
\usepackage{amsmath}
\usepackage{graphicx}
\usepackage{epsfig}
\usepackage{subfigure}
\usepackage{appendix}

\begin{document}

\title{Dynamic crystallization in a quantum Ising chain}
\author{K. L. Zhang}
\author{Z. Song}
\email{songtc@nankai.edu.cn}
\affiliation{School of Physics, Nankai University, Tianjin 300071, China}
\date{\today}

\begin{abstract}
The topological degeneracy of ground states in transverse field Ising chain
cannot be removed by local perturbation and allows it to be a promising
candidate for topological computation. We study the dynamic processes of
crystallization and dissolution for the gapped ground states in an Ising
chain. For this purpose, the real-space renormalization method is employed
to build an effective Hamiltonian that captures the low-energy physics of a
given system. We show that the ground state and the first-excited state of an $%
\left( N+1\right) $-site chain can be generated from that of the $N$-site
one by adding a spin adiabatically and vice versa. Numerical simulation
shows that the robust quasidegenerate ground states of finite-size chain
can be prepared with high fidelity from a set of noninteracting spins by a
quasiadiabatic process. As an application, we propose a scheme for
entanglement transfer between a pair of spins and two separable Ising chains
as macroscopic topological qubits.
\end{abstract}

\maketitle

\section{Introduction}

The transverse field Ising model \cite{pfeuty1970one} is a paradigm in both
traditional second-order quantum phase transition (QPT) based on spontaneous
symmetry breaking \cite{sachdev1999quantum} and topological QPT, which is immune to
local perturbation \cite{zhang2015topological, zhang2017majorana}. One of the most remarkable features of
the one-dimensional Ising chain is that its topological degenerate ground
states are protected from higher-energy excitations by the energy gap, and
are characterized by the existence of Majorana edge states, which are robust
to perturbations \cite{kitaev2001unpaired}. In the past few decades, the numerical
calculations \cite{botet1983finite, nightingale1986gap, white1993numerical, white2008spectral, shim2010artificial} and experimental investigation \cite{buyers1986experimental, morra1988spin, vcivzmar2008magnetic, delgado2013local} in quasi-one-dimensional complex compounds triggered the study of macroscopic quantum phenomena in quantum spin systems \cite{ruegg2003bose, ronnow2005quantum}. It has been reported that an interacting Ising spin chain can be simulated
by using Mott insulator spinless bosons in a tilted optical lattice \cite%
{simon2011quantum}. And the ground states of spin-1 antiferromagnetic
Heisenberg chain, which possesses a topological phase of matter known as the
Haldane phase \cite{haldane1983nonlinear, affleck1988valence, affleck1989quantum}, are also experimentally achievable \cite{hagiwara1990observation, glarum1991observation, xu2018realizing}. A promising application of such quantum spin systems is
physical implementation of quantum information processing devices based on a 
solid-state system \cite{petta2005coherent, korkusinski2009coded, johnson2011quantum, hsieh2012physics, sarma2015majorana}. Thanks to the intrinsic stability of the topological feature, a system
with topological phase can be a promising platform for quantum computation
and information processing \cite{nayak2008non, stern2010non, alicea2012new}. It motivates us to
develop an alternative candidate for a macroscopic qubit based on the Ising
chain in the topological phase, due to the fact that its degenerate ground
states are robust against the disordered perturbation.

In this work, we explore a way to prepare the ground state and the first-excited
state of an Ising chain on demand by dynamic process of crystallization,
generating robust macroscopic quantum states against disordered
perturbation. Based on an analytical perturbation analysis, we employ a
real-space renormalization method to build an effective Hamiltonian that
captures the low-energy physics of a given system. Within the topological
nontrivial region, the effective Hamiltonian for an $\left( N+1\right) $%
-site chain is obtained from the ground state and first-excited state of an $N$-site
chain. It is a modified two-spin Ising model, which is exactly solvable and
allows one to design an adiabatic passage for crystallization or
dissolution. We show that the ground state and the first-excited state of an $\left(
N+1\right) $-site chain can be generated from that of the $N$-site one by
adding a spin adiabatically and vice versa. Starting from $N=1$, as a seed
crystal, numerical simulation is performed for the quasiadiabatic process,
confirming our prediction. As an application in quantum information
processing, we demonstrate the scheme of entanglement transfer between a
pair of qubits and two Ising chains, as macroscopic objects.

This paper is organized as follows. In Sec. \ref{Model and symmetry}, we
present the model and its symmetries. In Sec. \ref{Effective Hamiltonian},
an effective Hamiltonian that captures the low-energy physics is obtained
based on a real-space renormalization method. In Sec. \ref{Dynamic
crystallization}, we propose an adiabatic passage for crystallization, which
generates the ground state and the first-excited state of a finite-size chain from
simple spin configurations. It allows the scheme of entanglement transfer
from two qubits to two Ising chains, creating a macroscopic entangled state.
In Sec. \ref{Quenched disordered perturbation and entanglement distillation}%
, we investigate the robustness of the ground states in the presence of
quenched disordered perturbation. We also propose an adiabatic passage for
entanglement distillation from an obtained macroscopic entangled state.
Section \ref{Discussion} summarizes the results and explores their
implications.

\section{Model and symmetries}

\label{Model and symmetry}

We consider the Hamiltonian of a transverse field Ising model 
\begin{equation}
H=\sum_{j=1}^{N-1}J_{j}\sigma _{j}^{x}\sigma
_{j+1}^{x}+\sum_{j=1}^{N}g_{j}\sigma _{j}^{z},  \label{H_Ising}
\end{equation}%
on a chain with open boundary\ condition, where $\sigma _{j}^{\alpha }$ ($%
\alpha =x,$ $y,$ $z$) are the Pauli operators on site $j$ and parameters $%
J_{j}$ and $g_{j}$ are position and time dependent, without losing the
generality. The second term and quantity $\sigma ^{z}=\sum_{j=1}^{N}\sigma
_{j}^{z}$ have common eigenstates, and the first term breaks the
conservation of this quantity. However, the parity of the eigenvalue of $%
\sigma ^{z}$\ is conservative, i.e., we always have%
\begin{equation}
\left[ p,H\right] =0,
\end{equation}%
where the parity operator%
\begin{equation}
p=\prod_{j=1}^{N}(-\sigma _{j}^{z}).
\end{equation}%
We start from the simple case with uniform parameters, $J_{j}=J$ and $%
g_{j}=g $, to analyze the property of the ground state. For finite $N$, the
parity of ground state with nonzero $g$ can be determined by the ground
state in the $g=\pm \infty $\ limit. It is due to the fact of nondegeneracy of
the ground state for finite $N$ and nonzero $g$. Actually, it has been shown
that the spectrum of $H$\ can be constructed based on the positive levels of
the corresponding Su-Schrieffer-Heeger (SSH) chain \cite{kitaev2001unpaired, su1979solitons}. The
nonzero energy levels of an SSH chain result in the fact that the ground-state energy is nondegenerate except at $g=0$. Here we give the conclusions of the parity
of ground state: (i) for even $N$, we have $p=1$ for the ground state with
any $g\neq 0 $, while (ii) for odd $N$, we have $p=\mathrm{sgn}\left(
g\right) $.

For a finite $N$ system with a periodic boundary condition, the exact solution
can be obtained \cite{pfeuty1970one} and the ground state obeys the same rule. It
is the common sense that the property of the model is not sensitive to the
boundary condition in the thermodynamic limit. Nevertheless, here we would like
to emphasize that the Hamiltonian with infinite $N$ possesses an exclusive symmetry in the topological nontrivial region $0<\left\vert g/J\right\vert <1$. It can be checked that there
exists a nonlocal spin operator (see the Appendix) 
\begin{eqnarray}
D_{N} &=&\frac{1}{2}\sqrt{1-\left( \frac{g}{J}\right) ^{2}}%
\sum_{j=1}^{N}\prod\limits_{l<j}\left( -\sigma _{l}^{z}\right)  \notag \\
&&\times \left[ \left( -\frac{g}{J}\right) ^{j-1}\sigma _{j}^{x}-i\left( -%
\frac{g}{J}\right) ^{N-j}\sigma _{j}^{y}\right] ,  \label{D_N}
\end{eqnarray}%
satisfying the commutation relations%
\begin{equation}
\lbrack D_{N},H]=[D_{N}^{\dag },H]=0,  \label{symmetry}
\end{equation}%
\begin{equation}
\{D_{N},D_{N}^{\dag }\}=1,(D_{N})^{2}=(D_{N}^{\dag })^{2}=0,  \label{commu_D}
\end{equation}%
Besides the method presented in the Appendix, $D_{N}$
can also be obtained by using the iterative method presented in Ref. \cite{fendley2016strong}, in which $\Psi$ is a Majorana operator, while $D_{N}$ is a fermion
operator satisfying Eq. (\ref{commu_D}). The term $\prod\nolimits_{l<j}%
\left( -\sigma _{l}^{z}\right)$ in $D_{N}$ is similar to the $%
\nu_{l} $ operators in Ref. \cite{katsura1962statistical}, and they both arise from the
transformation between spin operators and fermion operators. We would like
to point out that the commutation relation in Eq. (\ref{symmetry}) can be
regarded as a symmetry of the system. Importantly, such a symmetry is
conditional, requiring $0<\left\vert g/J\right\vert <1$, large $N$ limit, and
open boundary condition. The first two conditions are in accord with the symmetry
breaking mechanism for QPT \cite{sachdev1999quantum}. The transverse
field Ising model with periodic boundary condition \cite{pfeuty1970one} breaks the
symmetry Eq. (\ref{symmetry}). This can be related to the fact that the SSH
chain supports two zero-eigenenergy edge states in the topological
nontrivial region (see the Appendix), while the SSH ring does not support these
two modes.

Furthermore, the commutation relation in Eq. (\ref{symmetry}) guarantees the
existence of degeneracy of the eigenstates. There might be a set of
degenerate eigenstates $\left\{ \left\vert \psi _{n}^{+}\right\rangle
,\left\vert \psi _{n}^{-}\right\rangle \right\} $ of $H$\ with eigenenergy $%
E_{n}$, in two invariant subspaces, i.e.,%
\begin{equation}
H\left\vert \psi _{n}^{\pm }\right\rangle =E_{n}\left\vert \psi _{n}^{\pm
}\right\rangle
\end{equation}%
and%
\begin{equation}
p\left\vert \psi _{n}^{\pm }\right\rangle =\pm \left\vert \psi _{n}^{\pm
}\right\rangle .
\end{equation}%
Importantly, we have the relations%
\begin{eqnarray}
D_{N}\left\vert \psi _{n}^{+}\right\rangle &=&\left\vert \psi
_{n}^{-}\right\rangle ,D_{N}^{\dag }\left\vert \psi _{n}^{-}\right\rangle
=\left\vert \psi _{n}^{+}\right\rangle ,  \notag \\
D_{N}^{\dag }\left\vert \psi _{n}^{+}\right\rangle &=&D_{N}\left\vert \psi
_{n}^{-}\right\rangle =0.  \label{mapping}
\end{eqnarray}%
Especially, applying the operator $D_{N}$ ($D_{N}^{\dag }$) on the lowest-energy eigenstates $\left\vert \psi _{\mathrm{g}}^{+}\right\rangle $\ and $%
\left\vert \psi _{\mathrm{g}}^{-}\right\rangle $ of $H$\ in two invariant
subspaces, we have%
\begin{eqnarray}
D_{N}\left\vert \psi _{\mathrm{g}}^{+}\right\rangle &=&\left\vert \psi _{%
\mathrm{g}}^{-}\right\rangle ,D_{N}^{\dag }\left\vert \psi _{\mathrm{g}%
}^{-}\right\rangle =\left\vert \psi _{\mathrm{g}}^{+}\right\rangle ,  \notag
\\
D_{N}^{\dag }\left\vert \psi _{\mathrm{g}}^{+}\right\rangle
&=&D_{N}\left\vert \psi _{\mathrm{g}}^{-}\right\rangle =0,
\end{eqnarray}%
and then $\left\vert \psi _{\mathrm{g}}^{+}\right\rangle $\ and $\left\vert
\psi _{\mathrm{g}}^{-}\right\rangle $ are degenerate ground states. In the
rest of this paper, we denote eigenstates $\left\vert \psi _{\mathrm{g}%
}^{+}\right\rangle $\ and $\left\vert \psi _{\mathrm{g}}^{-}\right\rangle $
by $\left\vert \psi _{\mathrm{g}}^{N}\right\rangle $\ and $\left\vert \psi _{%
\mathrm{e}}^{N}\right\rangle $, representing the ground state and the first-excited
state, respectively.

Such a symmetry is also responsible for the topological degeneracy. In fact,
there also exists an operator $D_{N}$ for the Hamiltonian in the presence
of slight disordered deviations on the uniform $J_{j}=J$ and $g_{j}=g$,
since the edge modes of the SSH chain are robust against disordered perturbation
(see the Appendix). Thus the degeneracy of ground states cannot be lifted by
local perturbation. It is desirable to employ such two states as two
orthonormal basis states of a topological qubit. To this end, a basic task
is to find a way to prepare the ground state and the first-excited state (which are
quasidegenerate) of the Ising chain.

\begin{figure}[tbh]
\centering
\includegraphics[width=0.48\textwidth]{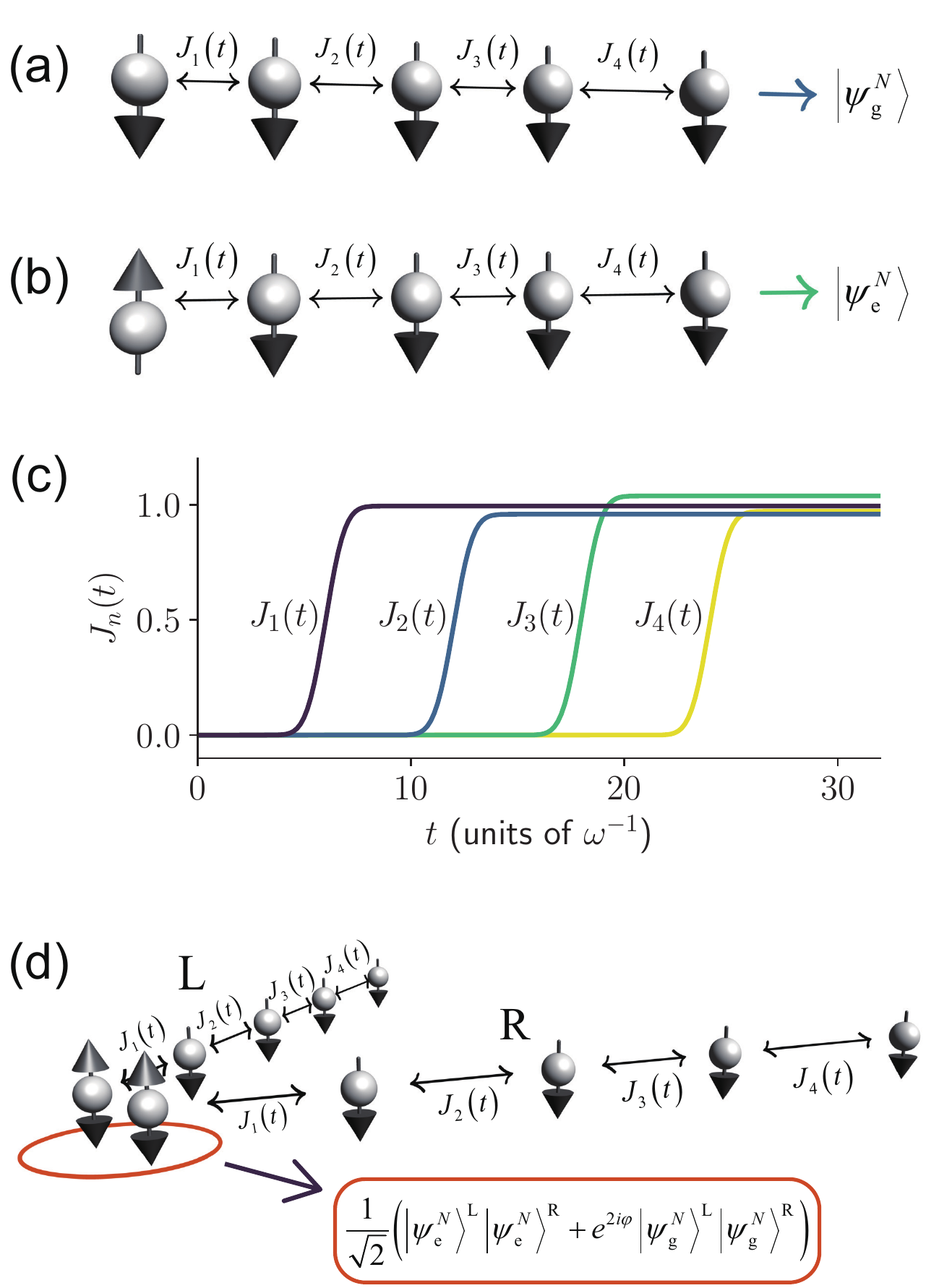}
\caption{Panels (a) and (b) are the schematic illustrations of the dynamic
crystallization process. The spin configurations in panels (a) and (b) are
adiabatically evolving into the ground state and the first-excited state,
respectively, by turning on the couplings $\left\{J_{n}(t)\right\} $ one by
one adiabatically. (c) Plot of the couplings in the form of Eq. (\protect\ref%
{erf_fun}) as functions of time $t$. Here we take $\protect\omega =0.001$, $%
\protect\tau =6 $, and $\protect\lambda _{n}\approx 1$ as an example. (d)
Schematic illustration of the entanglement transfer process between a pair
of spins and two separable Ising chains.}
\label{fig1}
\end{figure}

\section{Effective Hamiltonian}

\label{Effective Hamiltonian}

In this section, we aim to establish the connection between a single spin
and an Ising chain. We will provide a way to generate the ground state and the 
first-excited state of an $(N+1)$-site Ising chain from that of the $N$%
-site one. To proceed, we consider a system which consists of two parts,
an Ising chain and a single spin. The Hamiltonian has the form 
\begin{eqnarray}
&&H^{(N+1)}=H_{0}+H^{\prime },  \label{H(N+1)} \\
&&H_{0}=J\sum_{i=1}^{N-1}\sigma _{i}^{x}\sigma
_{i+1}^{x}+g\sum_{i=1}^{N}\sigma _{i}^{z}+g\sigma _{N+1}^{z},
\end{eqnarray}%
with $0<g\ll J$, and the coupling between them is Ising type%
\begin{equation}
H^{\prime }=\lambda \sigma _{N}^{x}\sigma _{N+1}^{x},
\end{equation}%
with positive parameter $\lambda \ll J$. Here the $N$-site Ising chain $%
H_{0}-g\sigma _{N+1}^{z}$\ has gapped low-lying eigenstates $\left\vert \psi
_{\mathrm{e}}^{N}\right\rangle $ and $\left\vert \psi _{\mathrm{g}%
}^{N}\right\rangle $ with energy $E_{\mathrm{e}}^{(N)}$ and $E_{\mathrm{g}%
}^{(N)},$\ respectively, and the gap is sufficiently large, so that $%
H^{\prime }=$ $\lambda \left( \sigma _{N}^{+}+\sigma _{N}^{-}\right) \left(
\sigma _{N+1}^{+}+\sigma _{N+1}^{-}\right) $\ can be regarded as
perturbation. By adiabatically eliminating the excited levels, we obtain an
effective Hamiltonian 
\begin{equation}
H_{\mathrm{eff}}^{\left( N+1\right) }=J_{\mathrm{eff}}^{(N)}\sigma
_{0}^{x}\sigma _{N+1}^{x}+g_{\mathrm{eff}}^{(N)}\sigma _{0}^{z}+g\sigma
_{N+1}^{z}+\frac{E_{\mathrm{e}}^{(N)}+E_{\mathrm{g}}^{(N)}}{2},
\end{equation}%
where the Pauli matrices $\sigma _{0}^{x}\ $and $\sigma _{0}^{z}\ $are
defined as 
\begin{eqnarray}
\sigma _{0}^{x}\left\vert \psi _{\mathrm{e}}^{N}\right\rangle &=&\left\vert
\psi _{\mathrm{g}}^{N}\right\rangle ,\sigma _{0}^{x}\left\vert \psi _{%
\mathrm{g}}^{N}\right\rangle =\left\vert \psi _{\mathrm{e}}^{N}\right\rangle
,  \notag \\
\sigma _{0}^{z}\left\vert \psi _{\mathrm{e}}^{N}\right\rangle &=&\left\vert
\psi _{\mathrm{e}}^{N}\right\rangle ,\sigma _{0}^{z}\left\vert \psi _{%
\mathrm{g}}^{N}\right\rangle =-\left\vert \psi _{\mathrm{g}%
}^{N}\right\rangle .
\end{eqnarray}%
And the effective coupling and field are%
\begin{eqnarray}
J_{\mathrm{eff}}^{(N)} &=&\lambda \left\langle \psi _{\mathrm{g}%
}^{N}\right\vert \left( \sigma _{N}^{+}+\sigma _{N}^{-}\right) \left\vert
\psi _{\mathrm{e}}^{N}\right\rangle ,  \notag \\
g_{\mathrm{eff}}^{(N)} &=&\frac{E_{\mathrm{e}}^{(N)}-E_{\mathrm{g}}^{(N)}}{2}%
.
\end{eqnarray}%
The effective Hamiltonian $H_{\mathrm{eff}}^{(N+1)}$ is a modified two-site
Ising model\ if $g_{\mathrm{eff}}^{(N)}\neq g$, describing low-lying
eigenstates of the original Hamiltonian $H^{(N+1)}$ in Eq. (\ref{H(N+1)}).

We readily obtain the ground state and the first-excited state of the effective
Hamiltonian $H_{\mathrm{eff}}^{\left( N+1\right) }$, which are%
\begin{equation}
\left\vert \psi _{\mathrm{eff}}^{\mathrm{g}}\right\rangle =\frac{1}{\sqrt{%
1+\left( \xi _{N}^{+}\right) ^{2}}}\left( \left\vert \psi _{\mathrm{g}%
}^{N}\right\rangle \left\vert \downarrow \right\rangle _{N+1}-\xi
_{N}^{+}\left\vert \psi _{\mathrm{e}}^{N}\right\rangle \left\vert \uparrow
\right\rangle _{N+1}\right)
\end{equation}%
and 
\begin{equation}
\left\vert \psi _{\mathrm{eff}}^{\mathrm{e}}\right\rangle =\frac{1}{\sqrt{%
1+\left( \xi _{N}^{-}\right) ^{2}}}\left( \left\vert \psi _{\mathrm{e}%
}^{N}\right\rangle \left\vert \downarrow \right\rangle _{N+1}-\xi
_{N}^{-}\left\vert \psi _{\mathrm{g}}^{N}\right\rangle \left\vert \uparrow
\right\rangle _{N+1}\right) ,
\end{equation}%
with energies 
\begin{equation}
E_{\mathrm{eff}}^{\mathrm{g}}=\frac{E_{\mathrm{e}}^{(N)}+E_{\mathrm{g}}^{(N)}%
}{2}-\Lambda _{N}^{+}
\end{equation}%
and 
\begin{equation}
E_{\mathrm{eff}}^{\mathrm{e}}=\frac{E_{\mathrm{e}}^{(N)}+E_{\mathrm{g}}^{(N)}%
}{2}-\Lambda _{N}^{-},
\end{equation}%
respectively, where the coefficients 
\begin{eqnarray}
\xi _{N}^{\pm } &=&\frac{J_{\mathrm{eff}}^{(N)}}{g\pm g_{\mathrm{eff}%
}^{(N)}+\Lambda _{N}^{\pm }},  \notag \\
\Lambda _{N}^{\pm } &=&\sqrt{\left[ g\pm g_{\mathrm{eff}}^{(N)}\right] ^{2}+%
\left[ J_{\mathrm{eff}}^{(N)}\right] ^{2}}.
\end{eqnarray}%
We note that in the case of 
\begin{equation}
g\pm g_{\mathrm{eff}}^{(N)}>0,  \label{g condition}
\end{equation}%
the ground state and the first-excited state reduce to $\left\vert \psi _{\mathrm{g}%
}^{N}\right\rangle \left\vert \downarrow \right\rangle _{N+1}$ and $%
\left\vert \psi _{\mathrm{e}}^{N}\right\rangle \left\vert \downarrow
\right\rangle _{N+1}$, respectively, when $J_{\mathrm{eff}}^{(N)}=0$. This is
crucial for the present work. A straightforward conclusion is that the
ground state and the first-excited state $\left\vert \psi _{\mathrm{g}%
}^{N+1}\right\rangle $\ and $\left\vert \psi _{\mathrm{e}}^{N+1}\right%
\rangle $ of\ an $\left( N+1\right) $-site chain can be obtained by adding a
down spin $\left\vert \downarrow \right\rangle _{N+1}$, and then
adiabatically increasing $\lambda $\ from $0$ to $J$. The explicit
expressions are%
\begin{eqnarray}
\mathcal{U}\left( H^{\prime }\right) \left\vert \psi _{\mathrm{g}%
}^{N}\right\rangle \left\vert \downarrow \right\rangle _{N+1} &=&\left\vert
\psi _{\mathrm{g}}^{N+1}\right\rangle , \\
\mathcal{U}\left( H^{\prime }\right) \left\vert \psi _{\mathrm{e}%
}^{N}\right\rangle \left\vert \downarrow \right\rangle _{N+1} &=&\left\vert
\psi _{\mathrm{e}}^{N+1}\right\rangle ,
\end{eqnarray}%
where 
\begin{equation}
\mathcal{U}\left( H^{\prime }\right) =\mathcal{T}\exp \left[
-i\int_{0}^{\infty }H^{\prime }(t)\mathrm{d}t\right] ,
\end{equation}%
is the propagator of the Hamiltonian $H^{\prime }(t)=\lambda (t)\sigma
_{N}^{x}\sigma _{N+1}^{x}$, with $\lambda (t)$ being a very slow function
as an adiabatic passage, fulfilling $\lambda (0)=0$ and $\lambda (\infty )=J 
$. Here, $\mathcal{T}$ is the time-ordering operator.  The valid maximum value of $d\lambda (t)/dt$ depends on the energy gap between two low-lying states and the higher excited states. Then due to the protection of the energy gap, the higher excitations are not affected.

\begin{figure*}[tbh]
\centering
\includegraphics[width=1\textwidth]{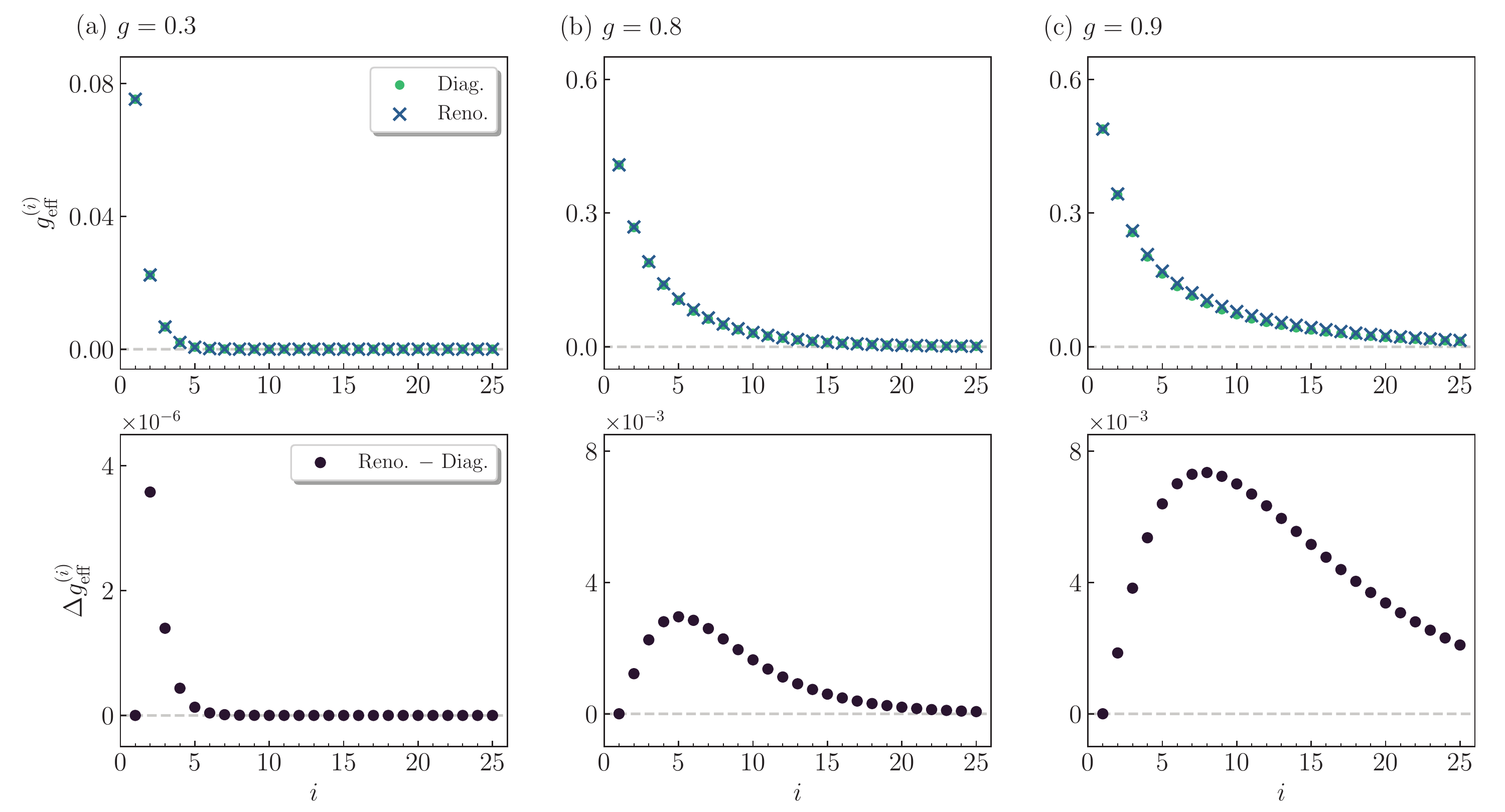}
\caption{Comparison between the strength of effective field $g_{\mathrm{eff}%
}^{(i)}$ in Eq. (\protect\ref{g(i+1)}) obtained from the renormalization
method and the exact diagonalization method. The upper panel is the results obtained from the two methods and the lower panel shows the difference between them. The parameters are $g_{i}=g=0.3$%
, $0.8$ and $0.9$ ($i\neq 1$) for panels (a), (b) and (c), respectively.
Other parameters are $J _{i}=\protect\lambda =1$ and $g_{1}=0.9g$. We can
see that, for the case of $g\ll 1$ and large $i$ ($N$), the results obtained from the renormalization method have relatively small errors. }
\label{fig2}
\end{figure*}

\section{Dynamic crystallization}

\label{Dynamic crystallization}

\begin{figure}[tbh]
\centering
\includegraphics[width=0.45\textwidth]{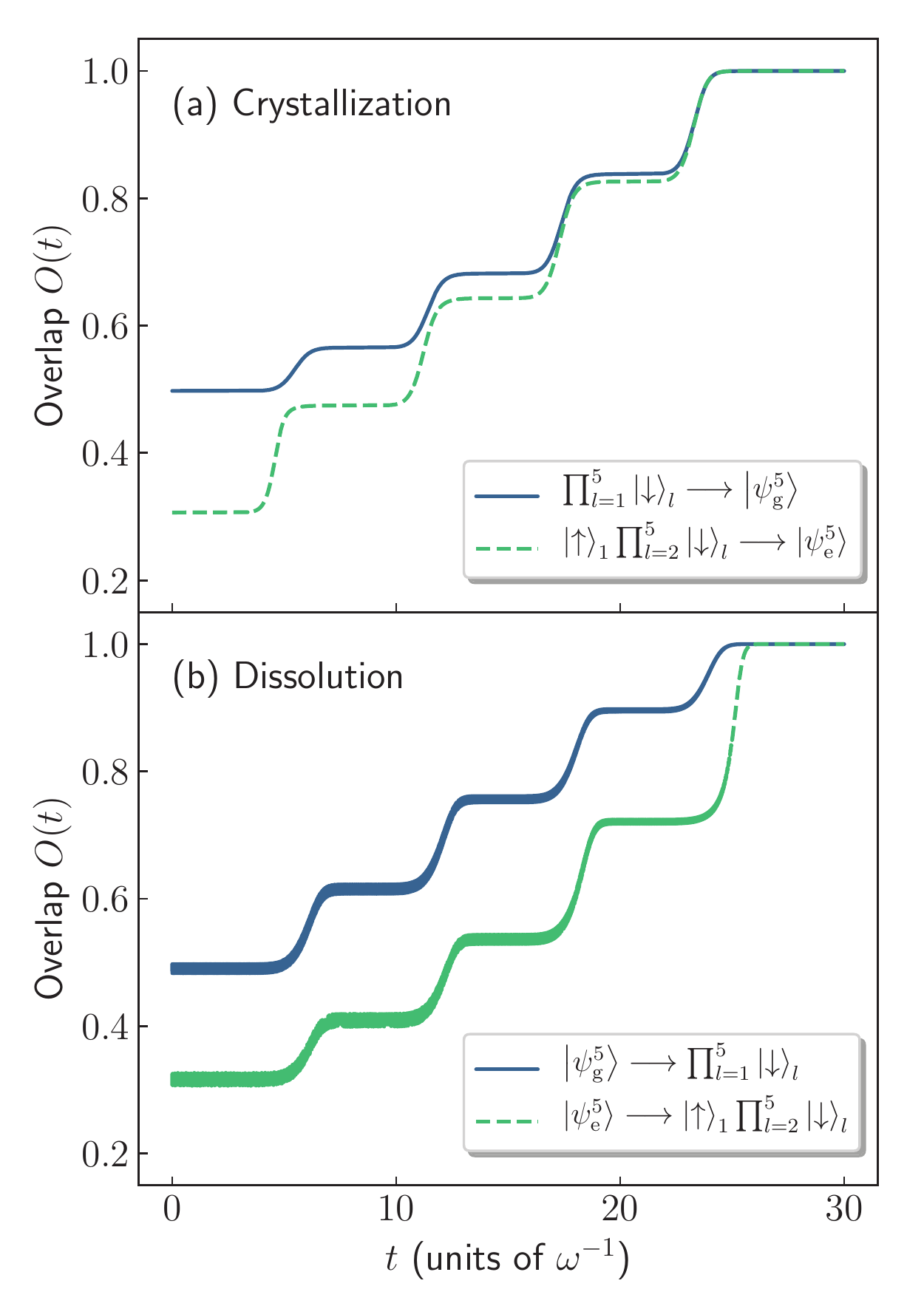}
\caption{Numerical results of the overlaps between target states and evolved
states for a five-site Ising chain. In the legends, the initial (target)
states are on the left (right) side of the arrows. (a) The dynamic
crystallization process. The ground (first-excited) state is generated
through numerical diagonalization for a uniform Ising chain with parameters $%
J=1$, $g=0.4$, and $N=5$. The time-dependent Hamiltonian is taken as the form
in Eq. (\protect\ref{H_DC}) with the couplings $\left\{ J_{n}(t)\right\}$ in
the form of Eq. (\protect\ref{erf_fun}). The parameters are $g_{i}=g$ ($%
i\neq 1$), $g_{1}=0.9g$, $\protect\omega =0.001$, $\protect\tau=6$, and $%
\protect\lambda_{n}=1$. (b) The dynamic dissolution process, which can be
regarded as a time reversal of the crystallization process. In contrast to
the crystallization process, here the initial state is taken as the ground
(first-excited) state of the Ising chain, and the couplings $\left\{
J_{n}(t)\right\} $ are removed one by one adiabatically.}
\label{fig3}
\end{figure}

So far, we have shown how to map the ground state and the first-excited state of an $%
N $-site Ising chain to that of the $\left( N+1\right) $-site system. It may
provide a way to generate the ground state and the first-excited state of finite-size
chain from simple spin configurations by the process of dynamic
crystallization. Here, we refer to this process as "crystallization" due to
the following reasons: (i) the process is about the noninteracting
particles evolving to the coupled array and (ii) the final state is determined by
the initial state of the first site as a seed crystal. The schematic
illustration of this process is shown in Figs. \ref{fig1}(a) and \ref{fig1}(b) for a five-site system.

We describe such a process by a time-dependent Hamiltonian 
\begin{equation}
H_{\mathrm{DC}}(t)=\sum_{i=1}^{N-1}J_{i}(t)\sigma _{i}^{x}\sigma
_{i+1}^{x}+\sum_{i=1}^{N}g_{i}\sigma _{i}^{z},  \label{H_DC}
\end{equation}%
where $\left\{ J_{n}(t)\right\} $\ is a series of slow functions 
fulfilling $J_{n} (0)=0$ and $J_{n} (\infty )=\mathrm{const.}$ to switch the
couplings along the chain one by one consecutively and quasiadiabatically.
A typical form of $\left\{ J_{n}(t)\right\} $ is the error function 
\begin{equation}
J_{n}(t)=\frac{\lambda _{n}}{2}\left\{ \mathrm{erf}\left[ \omega \left(
t-n\tau \right) \right] +1\right\} ,  \label{erf_fun}
\end{equation}%
where $\lambda _{n}$ is the strength of the final coupling. The shape of
function $J_{n}(t)$ is plotted in Fig. \ref{fig1}(c). In the following, we
estimate the possible result, based on the renormalization method. The basic
idea is as follows. According to the analysis in the previous section, in
the adiabatic regime, the dynamics in the duration of switching on $J_{i}$
is governed approximately by the effective Hamiltonian in the form 
\begin{equation}
H_{\mathrm{eff}}^{(i+1)}=J_{\mathrm{eff}}^{(i)}\sigma _{0}^{x}\sigma
_{i+1}^{x}+g_{\mathrm{eff}}^{(i)}\sigma _{0}^{z}+g_{i+1}\sigma _{i+1}^{z}+%
\frac{E_{\mathrm{eff}}^{\mathrm{e}(i)}+E_{\mathrm{eff}}^{\mathrm{g}(i)}}{2},
\end{equation}%
where Pauli operators $\sigma _{0}^{x}$ and $\sigma _{0}^{z}$ take actions
on the ground state and the first-excited state of the $i$-site chain; parameters $J_{%
\mathrm{eff}}^{(i)}$, $g_{\mathrm{eff}}^{(i)}$, $E_{\mathrm{eff}}^{\mathrm{e}%
(i)}$, and $E_{\mathrm{eff}}^{\mathrm{g}(i)}$ are obtained from $H_{\mathrm{%
eff}}^{(i)}$. Then $H_{\mathrm{eff}}^{(i+1)}$ generates the parameters in
the effective Hamiltonian $H_{\mathrm{eff}}^{(i+2)}$:\ 
\begin{eqnarray}
g_{\mathrm{eff}}^{(i+1)} &=&\frac{E_{\mathrm{eff}}^{\mathrm{e}(i+1)}-E_{%
\mathrm{eff}}^{\mathrm{g}(i+1)}}{2}  \notag \\
&=&\frac{1}{2}\left( \Lambda _{i}^{+}-\Lambda _{i}^{-}\right)  \label{g(i+1)}
\end{eqnarray}%
and%
\begin{eqnarray}
J_{\mathrm{eff}}^{(i+1)} &=&\lambda _{i}\left\langle \psi _{\mathrm{eff}}^{%
\mathrm{g}(i+1)}\right\vert \left( \sigma _{i+1}^{+}+\sigma
_{i+1}^{-}\right) \left\vert \psi _{\mathrm{eff}}^{\mathrm{e}%
(i+1)}\right\rangle  \notag \\
&=&-\frac{\lambda _{i}\left( \xi _{i}^{+}+\xi _{i}^{-}\right) }{\sqrt{%
1+\left( \xi _{i}^{+}\right) ^{2}}\sqrt{1+\left( \xi _{i}^{-}\right) ^{2}}}.
\label{J(i+1)}
\end{eqnarray}%
We start from $i=1$ with $J_{\mathrm{eff}}^{(1)}=\lambda _{1}$, $g_{\mathrm{%
eff}}^{(1)}=g_{1}$, and $E_{\mathrm{eff}}^{\mathrm{e}(1)}=-E_{\mathrm{eff}}^{%
\mathrm{g}(1)}=g_{1}$, i.e.,%
\begin{equation}
H_{\mathrm{eff}}^{(2)}=\lambda _{1}\sigma _{1}^{x}\sigma
_{2}^{x}+g_{1}\sigma _{1}^{z}+g_{2}\sigma _{2}^{z},
\end{equation}%
which is a two-site Ising model. Then effective parameters $g_{\mathrm{eff}%
}^{(i)}$ and $J_{\mathrm{eff}}^{(i)}$ with $i>1$ are obtained by the
iteration from Eqs. (\ref{g(i+1)}) and (\ref{J(i+1)}) or $H_{\mathrm{%
eff}}^{(i)}$. In order to verify the proposed approximate approach, we
compare the strength of effective field $g_{\mathrm{eff}}^{(i)}$ obtained
from the renormalization method in Eq. (\ref{g(i+1)}) and the exact
diagonalization method. The plots in Fig. \ref{fig2} show that  for small $g$ and large $i$ ($N$), the approximate results have relatively small errors.

Importantly, when a set of obtained parameters $\left\{ g_{\mathrm{eff}%
}^{(i)}\right\} $\ satisfies the condition 
\begin{equation}
g_{i+1}\pm g_{\mathrm{eff}}^{(i)}>0,
\end{equation}%
we have%
\begin{equation}
\mathcal{U}\left( H_{\mathrm{DC}}\right) \left( \alpha \left\vert \uparrow
\right\rangle _{1}+\beta \left\vert \downarrow \right\rangle _{1}\right)
\prod_{l=2}^{N}\left\vert \downarrow \right\rangle _{l}=\alpha \left\vert
\psi _{\mathrm{e}}^{N}\right\rangle +e^{i\varphi }\beta \left\vert \psi _{%
\mathrm{g}}^{N}\right\rangle ,  \label{crystallization}
\end{equation}%
where $\mathcal{U}\left( H_{\mathrm{DC}}\right) =\mathcal{T}\exp \left[
-i\int_{0}^{\infty }H_{\mathrm{DC}}(t)\mathrm{d}t\right] $, and $\varphi $\
is a dynamical phase. This process is similar to that of dynamic
crystallization in the case $\alpha =0$\ or $\beta =0$. Here the first spin
at state $\left\vert \uparrow \right\rangle _{1}(\left\vert \downarrow
\right\rangle _{1})$\ takes the role of a seed crystal, which determines the
state $\left\vert \psi _{\mathrm{e}}^{N}\right\rangle (\left\vert \psi _{%
\mathrm{g}}^{N}\right\rangle )$\ of the crystal with large $N$. We
demonstrate the dynamic crystallization by numerical simulation in a 
finite-size system. We use the overlap $O(t)=\left\vert \langle \Psi _{%
\mathrm{T}}\left\vert \Psi (t)\right\rangle \right\vert $ between the target
state $\left\vert \Psi _{\mathrm{T}}\right\rangle =\left\vert \psi _{\mathrm{%
g}}^{N}\right\rangle (\left\vert \psi _{\mathrm{e}}^{N}\right\rangle )$ obtained from numerical diagonalization and the evolved state%
\begin{equation}
\left\vert \Psi (t)\right\rangle =\mathcal{T}\exp \left[ -i\int_{0}^{t}H_{%
\mathrm{DC}}(t^{\prime })\mathrm{d}t^{\prime }\right] \left\vert \Psi
(0)\right\rangle ,
\end{equation}%
where $\left\vert \Psi (0)\right\rangle =\left\vert \downarrow \right\rangle
_{1}(\left\vert \uparrow \right\rangle _{1})\prod_{l=2}^{N}\left\vert
\downarrow \right\rangle _{l}$,\ to measure the efficiency of the process.
The overlaps $O(t)$\ are plotted in Fig. \ref{fig3}(a). Meanwhile, as a time
reversal of the crystallization process, the simulation for the dissolution
process is also performed [see Fig. \ref{fig3}(b)], where the corresponding
initial state and target state are $\left\vert \Psi (0)\right\rangle =\left\vert \psi _{\mathrm{g}%
}^{N}\right\rangle (\left\vert \psi _{\mathrm{e}}^{N}\right\rangle )$\ and $%
\left\vert \Psi _{\mathrm{T}}\right\rangle =\left\vert \downarrow
\right\rangle _{1}(\left\vert \uparrow \right\rangle
_{1})\prod_{l=2}^{N}\left\vert \downarrow \right\rangle _{l}$. Here the
computation is performed by using a uniform mesh in the time discretization
for the time-dependent Hamiltonian $H_{\mathrm{DC}}(t)$. The results are in accord
with our predictions for both processes.

We would like to point out that the dynamic crystallization process cannot
map a qubit state onto a many-qubit two-level state due to the uncertainty
of the phase $\varphi $. However, as an application in quantum information
processing, this makes it possible to realize the entanglement transfer from
a pair of qubits to two independent Ising chains, as macroscopic objects.
Entanglement is considered to be one of the most profound features of
quantum mechanics \cite{einstein1935can, bell1987speakable} and a very powerful resource for
quantum information processing and communication. Specifically, robust and
long-lived entanglement of material objects is a desirable task in quantum
information processing, including teleportation of quantum states of matter
and quantum memory \cite{horodecki2009quantum}. Here we propose a scheme to generate
robust entanglement between two quantum spin chains, as macroscopic objects.

The system we are concerned with is a simple extension of the original system,
described by the Hamiltonian

\begin{equation}
H_{\mathrm{D}}=H^{\mathrm{L}}+H^{\mathrm{R}},
\end{equation}%
where $H^{\mathrm{L}}\ $and $H^{\mathrm{R}}$\ represent two independent but
identical Ising chains described by Eq. (\ref{H_Ising}), respectively.
Consider an initial state, in which the first two spins are maximally
entangled, being state $\left( \left\vert \uparrow \right\rangle _{1}^{%
\mathrm{L}}\left\vert \uparrow \right\rangle _{1}^{\mathrm{R}}+\left\vert
\downarrow \right\rangle _{1}^{\mathrm{L}}\left\vert \downarrow
\right\rangle _{1}^{\mathrm{R}}\right) /\sqrt{2}$.\ Applying the process in
Eq. (\ref{crystallization}), we have 
\begin{eqnarray}
&&\frac{1}{\sqrt{2}}\mathcal{U}\left( H_{\mathrm{D}}\right) \left(
\left\vert \uparrow \right\rangle _{1}^{\mathrm{L}}\left\vert \uparrow
\right\rangle _{1}^{\mathrm{R}}+\left\vert \downarrow \right\rangle _{1}^{%
\mathrm{L}}\left\vert \downarrow \right\rangle _{1}^{\mathrm{R}}\right)
\prod_{l=2}^{N}\left\vert \downarrow \right\rangle _{l}^{\mathrm{L}%
}\prod_{l=2}^{N}\left\vert \downarrow \right\rangle _{l}^{\mathrm{R}}  \notag
\\
&&=\frac{1}{\sqrt{2}}\left( \left\vert \psi _{\mathrm{e}}^{N}\right\rangle ^{%
\mathrm{L}}\left\vert \psi _{\mathrm{e}}^{N}\right\rangle ^{\mathrm{R}%
}+e^{i2\varphi }\left\vert \psi _{\mathrm{g}}^{N}\right\rangle ^{\mathrm{L}%
}\left\vert \psi _{\mathrm{g}}^{N}\right\rangle ^{\mathrm{R}}\right) .
\label{entangle transfer}
\end{eqnarray}%
This represents entanglement transfer from two spins to two independent Ising
chains, keeping the maximal concurrence no matter what the value of $\varphi 
$. Such a process is schematically illustrated in Fig. \ref{fig1}(d).

\section{Quenched disordered perturbation and entanglement distillation}

\label{Quenched disordered perturbation and entanglement distillation}

In this section, we focus on the many-particle qubit in two aspects. (i) We
demonstrate the robustness of the macroscopic qubit state in the presence of
quenched disordered perturbation via a numerical simulation in finite
systems. (ii) We propose a scheme to distill the entanglement of two Ising
chains via a dissolution process, which transfers the entanglement from
chains to a fixed pair of spins.

\begin{figure*}[tbh]
\centering
\includegraphics[width=1\textwidth]{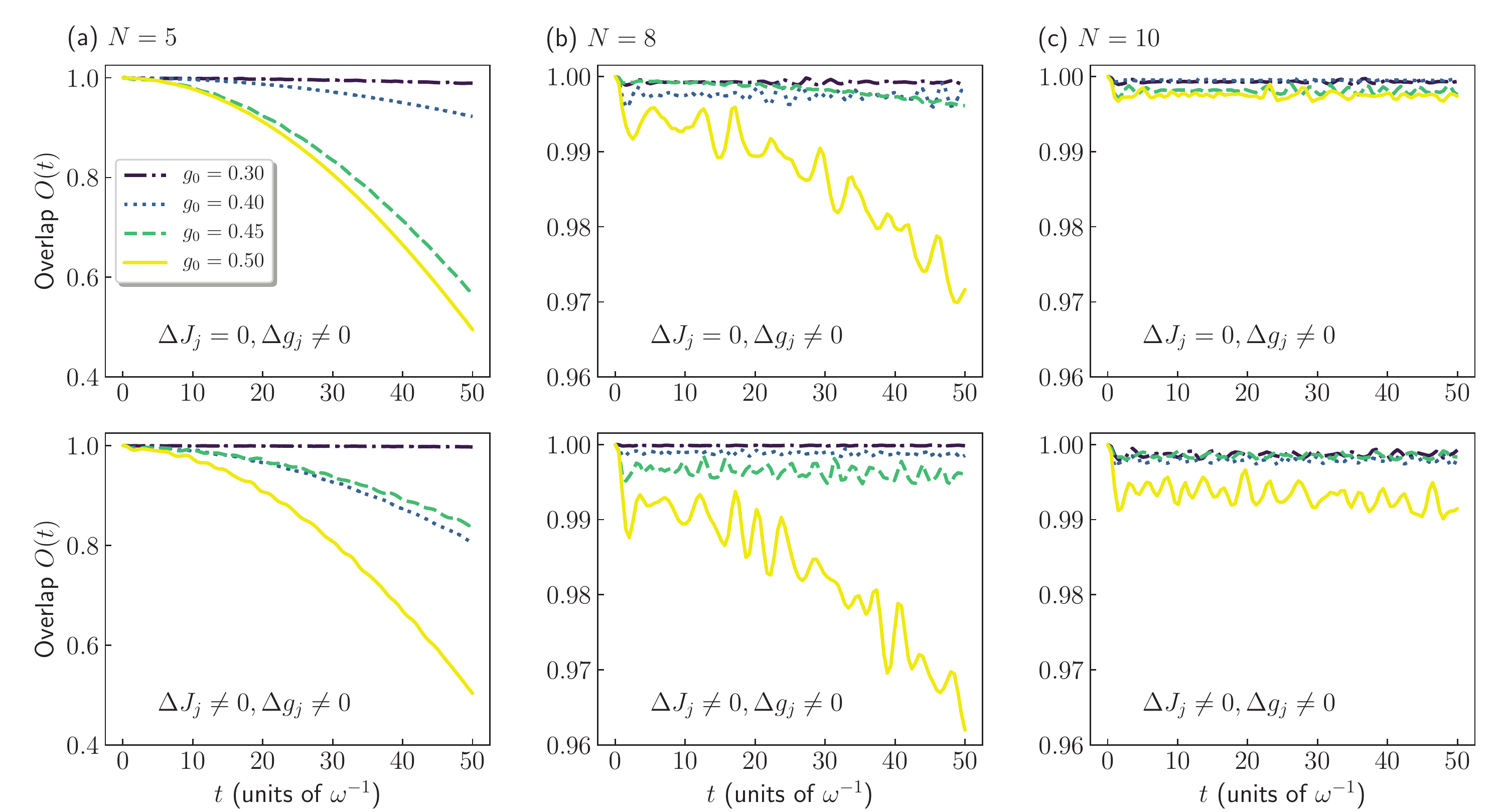}
\caption{Numerical results of the overlap between the initial state $%
\left\vert \Psi (0)\right\rangle$ and its evolved state $\left\vert \Psi
(t)\right\rangle$ under two kinds of quenched disordered perturbation in the forms of Eqs. (\ref{quench1}) and (\ref{quench2}) for different $N$ and $g_{0}$. The sizes of the chains are $N=5$, $N=8$ and $N=10$ for panels
(a), (b) and (c), respectively. The disorder strength is $R=0.2$ and the
parameters of the uniform Hamiltonian $H$ are $J_{j}=J_{0}=1$ and $g_{j}=g_{0}$.
Other parameters are $\protect\alpha =\protect\beta =1/\protect\sqrt{2}$ and 
$\protect\omega =1$.}
\label{fig4}
\end{figure*}

\subsection{Quenched disordered perturbation}

The advantage of the proposed many-particle qubit is that the quasidegenerate
ground states of an Ising chain are robust against local perturbation.
Technically speaking, there always exists a $D_{N}$ operator even when parameters 
$J_{j}$ and $g_{j}$ are slightly random (see the Appendix). This means that the
degeneracy cannot be lifted when the random perturbation is induced
adiabatically. Then during the process, $\alpha \left\vert \psi _{\mathrm{e}%
}^{N}(0)\right\rangle +\beta \left\vert \psi _{\mathrm{g}}^{N}(0)\right%
\rangle $ evolves to $\alpha \left\vert \psi _{\mathrm{e}}^{N}(t)\right%
\rangle +\beta \left\vert \psi _{\mathrm{g}}^{N}(t)\right\rangle $, without an 
extra time-dependent phase difference on $\alpha $ and $\beta $, keeping the
original quantum information. Here $\left\vert \psi _{\mathrm{g}%
}^{N}(t)\right\rangle $ and $\left\vert \psi _{\mathrm{e}}^{N}(t)\right%
\rangle $ are instantaneous ground state and first-excited state of the
time-dependent Hamiltonian. However, in practice, the appearance of
disordered perturbation from the environment is random in time. In the
following we consider an extreme case, in which a disordered perturbation\
is added as a quenching process, and investigate the effect of quenched
disordered perturbation on a many-spin qubit initial state $\left\vert \Psi
(0)\right\rangle =\alpha \left\vert \psi _{\mathrm{e}}^{N}(0)\right\rangle
+\beta \left\vert \psi _{\mathrm{g}}^{N}(0)\right\rangle $ by employing
numerical simulation for the time evolution on a finite $N$ system. We add a
time-dependent perturbation $H_{\mathrm{Ran}}$ to the uniform $N$-site Ising
chain. Here $H_{\mathrm{Ran}}$ takes the form%
\begin{equation}
H_{\mathrm{Ran}}=\sum_{j=1}^{N-1}\Delta J_{j}\sigma _{j}^{x}\sigma
_{j+1}^{x}+\sum_{j=1}^{N}\Delta g_{j}\sigma _{j}^{z},
\end{equation}%
in which the parameters take the Heaviside function\ of time.  We consider
the following two cases. (i) The coupling is homogeneous and the field is
random: 
\begin{eqnarray}
\Delta J_{j} &=&0, \notag\\
\Delta g_{j} &=&\frac{1}{2} g_{0}\Delta _{j}^{g}\left[ \mathrm{sgn}\left( t\right) +1%
\right].
\label{quench1}
\end{eqnarray}%
(ii) Both the coupling and the field are random:
\begin{eqnarray}
\Delta J_{j} &=& \frac{1}{2}J_{0}\Delta _{j}^{J}\left[ \mathrm{sgn}\left( t\right) +1%
\right], \notag\\
\Delta g_{j} &=& \frac{1}{2}g_{0}\Delta _{j}^{g}\left[ \mathrm{sgn}\left( t\right) +1%
\right].
\label{quench2}
\end{eqnarray}
Here $\{\Delta _{j}^{g},\Delta _{j}^{J}\}$ denotes a set of uniformly distributed random numbers within
the interval $(-R,R)$, taking the role of the disorder strength. We still
use the overlap $O(t)=\left\vert \langle \Psi (0)\left\vert \Psi
(t)\right\rangle \right\vert $\ between $\left\vert \Psi (0)\right\rangle $\
and the evolved state%
\begin{equation}
\left\vert \Psi (t)\right\rangle =\exp \left[ -i(H+H_{\mathrm{Ran}})t\right]
\left\vert \Psi (0)\right\rangle ,
\end{equation}%
to measure the influence of the quenched perturbation. The overlap $O(t)$\
for systems with different size and parameters are plotted in Fig. \ref{fig4}%
. The result with a fixed random strength $R$ shows that, for a fixed $N$%
, larger $g_{0}$ leads to smaller fidelity, while for a fixed $g_{0}$,
larger $N$ leads to larger fidelity. This indicates that, even for a finite-size
system with $N=10$, the ground state and the first-excited state are very robust for
the case with not large $g_{0}<0.5$.

\subsection{Entanglement distillation}

\begin{figure*}[tbh]
\centering
\includegraphics[width=0.8\textwidth]{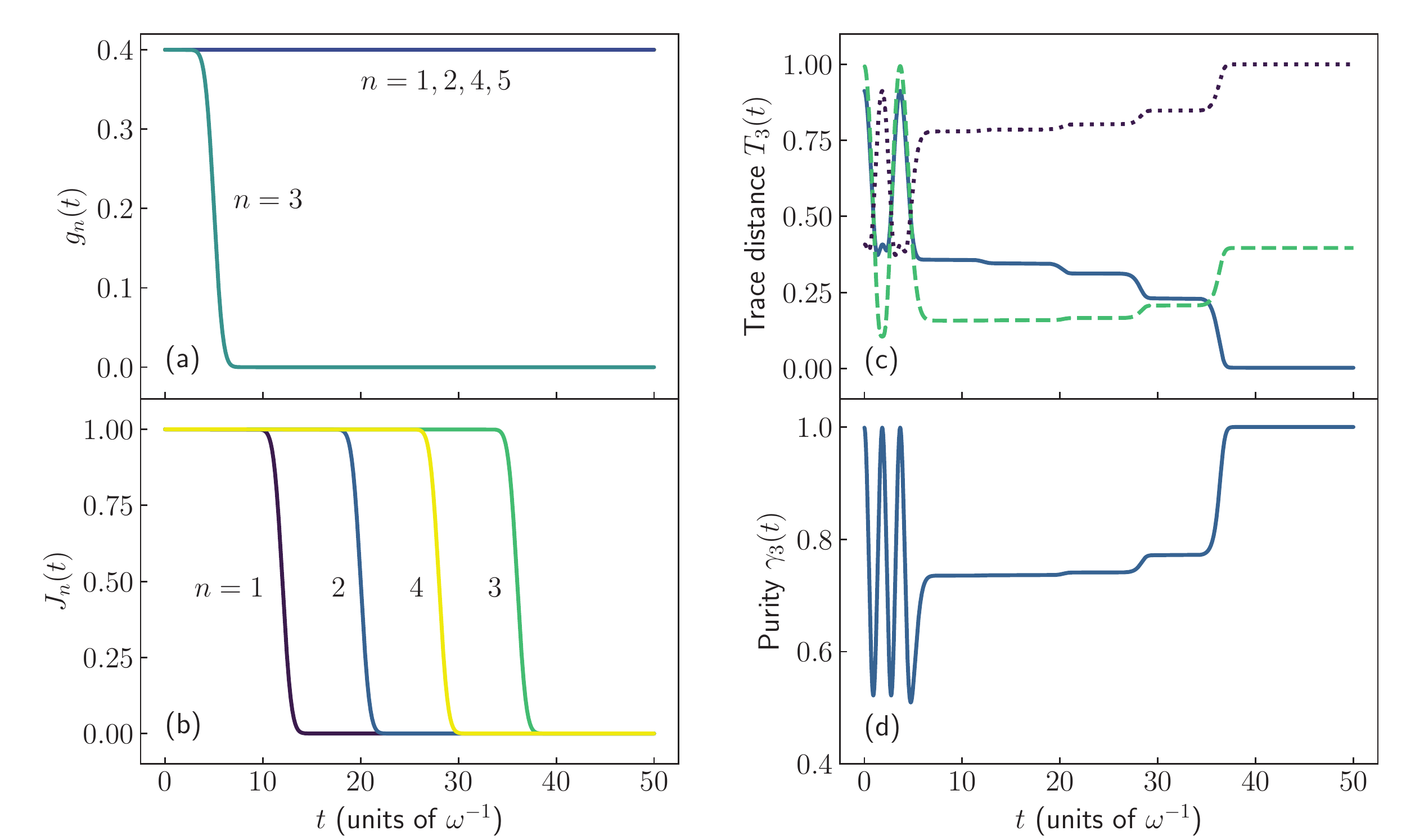}
\caption{(a) Adiabatic change of the strength of field $g_{n}(t)$. The
target spin at site $l=3$ is selected by deceasing the local field $g_{3}$
to $0$ adiabatically. (b) Adiabatic change of the strength of coupling $%
J_{n}(t)$. (c) The trace distance defined in Eq. (\protect\ref%
{trace_distance}) for different target qubit state $\overline{\protect\rho }%
_{l}$ at site $l=3$. The solid, dashed, and dotted lines represent the
results of target qubit states with different phase factors $\protect\varphi%
=-0.82i$, $0$, and $(\protect\pi-0.82i)$, respectively. (d) Purity of the
reduced density matrix $\protect\rho _{l}(t)$ for site $l=3$ defined in Eq. (%
\protect\ref{purity}). Other parameters are $N=5$, $\protect\alpha=\protect%
\beta=1/\protect\sqrt{2}$, and $\protect\omega =0.01 $ . }
\label{fig5}
\end{figure*}

In the following, we turn to demonstrate an adiabatic passage for
entanglement distillation from an obtained macroscopic entangled state. To
this end, we employ numerical simulation for the time evolution on finite $N$%
-site system $H$ in Eq. (\ref{H_Ising}). We start from an initial state $%
\left\vert \Psi (0)\right\rangle =\alpha \left\vert \psi _{\mathrm{e}%
}^{N}(0)\right\rangle +\beta \left\vert \psi _{\mathrm{g}}^{N}(0)\right%
\rangle $. At first step, an arbitrary target spin at site $l$\ is selected
by decreasing the local field $g_{l}$\ to zero, adiabatically. At second
step, after $g_{l}$\ vanishing, we remove the coupling $J_{j}$\ one by one,
adiabatically. The order of the dissolution is $J_{1}\rightarrow 0$, $%
J_{2}\rightarrow 0$,..., $J_{l-1}\rightarrow 0$, and then $J_{N-1}\rightarrow
0 $, $J_{N-2}\rightarrow 0$,..., $J_{l}\rightarrow 0$. The above parameters
as functions of time are plotted in Figs. \ref{fig5}(a) and \ref{fig5}(b) for a five-site system. During this process, we monitor the evolved state of the spin
at site $l$, by its $2\times 2$\ reduced density matrix%
\begin{equation}
\rho _{l}\left( t\right) =\mathrm{Tr}_{(l)}\left[ \left\vert \Psi
(t)\right\rangle \left\langle \Psi (t)\right\vert \right] ,
\end{equation}%
where $\mathrm{Tr}_{(l)}\left[ ...\right] $\ denotes taking the trace over
all the rest of the freedom. For the initial state $\left\vert \Psi
(0)\right\rangle $, the target qubit state is $\alpha \left\vert \uparrow
\right\rangle _{l}+e^{i\varphi }\beta \left\vert \downarrow \right\rangle
_{l}$, which is equivalent to the density matrix $\overline{\rho }%
_{l}=\alpha \alpha ^{\ast }\left\vert \uparrow \right\rangle
_{l}\left\langle \uparrow \right\vert _{l}$ $+\beta \beta ^{\ast }\left\vert
\downarrow \right\rangle _{l}\left\langle \downarrow \right\vert _{l}$ $%
+e^{-i\varphi }\alpha \beta ^{\ast }\left\vert \uparrow \right\rangle
_{l}\left\langle \downarrow \right\vert _{l}$ $+e^{i\varphi }\beta \alpha
^{\ast }\left\vert \downarrow \right\rangle _{l}\left\langle \uparrow
\right\vert _{l}$. To describe the efficiency, we employ the trace distance 
\begin{equation}
T_{l}(t)=\frac{1}{2}\mathrm{Tr}\left[ \sqrt{\left[ \overline{\rho }_{l}-\rho
_{l}\left( t\right) \right] ^{2}}\right] ,  \label{trace_distance}
\end{equation}%
which is a measure of the distinguishability between the evolved state and the target
state. We note that the final state $\rho _{l}\left( t\right) $\ depends on
the adiabatic passage due to the extra dynamic phase $\varphi $. Therefore,
even though state $\rho _{l}\left( t\right) $\ does not meet $\overline{\rho 
}_{l}$, it can still be a pure state. We have known that this fact makes it
possible for entanglement transfer as mentioned in Eq. (\ref{crystallization}%
). Then it is important to measure the purity of the reduced density matrix $%
\rho _{l}\left( t\right) $, which is defined as 
\begin{equation}
\gamma _{l}(t)=\mathrm{Tr}\left[ \rho _{l}\left( t\right) ^{2}\right] .
\label{purity}
\end{equation}%
The numerical results plotted in Figs. \ref{fig5}(c) and \ref{fig5}(d) show the
trace distance $T_{l}(t)$ and purity $\gamma _{l}(t)$ as functions of
time, verifying our prediction, i.e., 
\begin{equation}
\mathcal{U}\left( H\right) \left( \alpha \left\vert \psi _{\mathrm{e}%
}^{N}\right\rangle +\beta \left\vert \psi _{\mathrm{g}}^{N}\right\rangle
\right) =\left( \alpha \left\vert \uparrow \right\rangle _{l}+e^{i\varphi
}\beta \left\vert \downarrow \right\rangle _{l}\right) \prod_{j\neq
l}^{N}\left\vert \downarrow \right\rangle _{j},  \label{dissolution}
\end{equation}%
which is crucial\ to the scheme of entanglement distillation in the
following.

Now we extend the result in Eq. (\ref{dissolution}) to the two-chain system $%
H_{\mathrm{D}}=H^{\mathrm{L}}+H^{\mathrm{R}}$. We focus on the entanglement
distillation of the two-spin system from two entangled Ising chains. For
simplicity, we consider the case with $\alpha =\beta =1/\sqrt{2}$. From Eq. (%
\ref{dissolution}) we have%
\begin{eqnarray}
&&\frac{1}{\sqrt{2}}\mathcal{U}\left( H^{\mathrm{L,R}}\right) \left(
\left\vert \psi _{\mathrm{e}}^{N}\right\rangle ^{\mathrm{L,R}}\pm \left\vert
\psi _{\mathrm{g}}^{N}\right\rangle ^{\mathrm{L,R}}\right)  \notag \\
&&=\frac{1}{\sqrt{2}}\left( \left\vert \uparrow \right\rangle _{l}^{\mathrm{%
L,R}}\pm e^{i\varphi }\left\vert \downarrow \right\rangle _{l}^{\mathrm{L,R}%
}\right) \prod_{j\neq l}^{N}\left\vert \downarrow \right\rangle _{j}^{%
\mathrm{L,R}},
\end{eqnarray}%
which results in%
\begin{eqnarray}
&&\frac{1}{\sqrt{2}}\mathcal{U}\left( H_{\mathrm{D}}\right) \left(
\left\vert \psi _{\mathrm{e}}^{N}\right\rangle ^{\mathrm{L}}\left\vert \psi
_{\mathrm{e}}^{N}\right\rangle ^{\mathrm{R}}+\left\vert \psi _{\mathrm{g}%
}^{N}\right\rangle ^{\mathrm{L}}\left\vert \psi _{\mathrm{g}%
}^{N}\right\rangle ^{\mathrm{R}}\right)  \notag \\
&&=\frac{1}{\sqrt{2}}\left( \left\vert \uparrow \right\rangle _{l}^{\mathrm{L%
}}\left\vert \uparrow \right\rangle _{l}^{\mathrm{R}}+e^{i2\varphi
}\left\vert \downarrow \right\rangle _{l}^{\mathrm{L}}\left\vert \downarrow
\right\rangle _{l}^{\mathrm{R}}\right) \prod_{j\neq l}^{N}\left\vert
\downarrow \right\rangle _{j}^{\mathrm{L}}\prod_{j\neq l}^{N}\left\vert
\downarrow \right\rangle _{j}^{\mathrm{R}},
\end{eqnarray}%
by direct derivation. Accordingly, it also provides a way to transfer the
maximal pair entanglement from location $i$ to $l$, 
\begin{eqnarray}
&&\frac{1}{\sqrt{2}}\left( \left\vert \uparrow \right\rangle _{i}^{\mathrm{L}%
}\left\vert \uparrow \right\rangle _{i}^{\mathrm{R}}+\left\vert \downarrow
\right\rangle _{i}^{\mathrm{L}}\left\vert \downarrow \right\rangle _{i}^{%
\mathrm{R}}\right)  \notag \\
&&\longrightarrow\frac{1}{\sqrt{2}}\left( \left\vert \uparrow \right\rangle
_{l}^{\mathrm{L}}\left\vert \uparrow \right\rangle _{l}^{\mathrm{R}%
}+e^{i\varphi ^{\prime }}\left\vert \downarrow \right\rangle _{l}^{\mathrm{L}%
}\left\vert \downarrow \right\rangle _{l}^{\mathrm{R}}\right) .
\end{eqnarray}

\section{Discussion}
\label{Discussion}
In this paper, we have studied the relation between the gapped
quasidegenerate ground states of an $N$-site Ising chain and that\ of the $%
\left( N+1\right) $-site chain, based on which the real-space
renormalization method is developed. It allows us to build the effective
Hamiltonian, which is an exactly solvable modified two-site Ising model and
captures the low-energy physics of a given system. Numerical calculation
shows that such an effective Hamiltonian has higher efficiency and is\ a
feasible method for a large-size system. Due to the protection of the energy gap,
this approximate description provides an alternative way to prepare the
ground state and the first-excited state of an Ising chain on demand by dynamic
processes of crystallization, generating robust macroscopic quantum states
against disordered perturbation. To demonstrate the potential application of
our finding, we proposed a scheme of entanglement transfer between a pair of
qubits and two Ising chains, as macroscopic topological qubits. Our work,
including the numerical result for a small size system, reveals that
transverse field Ising chains can be utilized for developing inherently
robust artificial devices for topological quantum information processing and
communication.

\acknowledgments This work was supported by the National Natural Science
Foundation of China (under Grant No. 11874225).

\appendix

\section*{Appendix}

\label{Appendix} \setcounter{equation}{0} \renewcommand{\theequation}{A%
\arabic{equation}} \renewcommand{\thesubsection}{\arabic{subsection}}

In this appendix, we will show the method of obtaining the nonlocal operator 
$D_{N}$ in Eq. (\ref{D_N}), as well as demonstrate its
robustness. This method can also be used to analyze a non-Hermitian model \cite{zhang2020ising}. 
Starting from the Ising chain $H$ in Eq. (\ref{H_Ising}) with uniform
parameters $J_{j}=J$ and $g_{j}=g$, we first perform the Jordan-Wigner
transformation \cite{jordan1993paulische} 
\begin{eqnarray}
\sigma _{j}^{x} &=&\prod\limits_{l<j}\left( 1-2c_{l}^{\dagger }c_{l}\right)
\left( c_{j}+c_{j}^{\dagger }\right) ,  \notag \\
\sigma _{j}^{y} &=&i\prod\limits_{l<j}\left( 1-2c_{l}^{\dagger }c_{l}\right)
\left( c_{j}-c_{j}^{\dagger }\right) ,  \notag \\
\sigma _{j}^{z} &=&2c_{j}^{\dag }c_{j}-1,
\end{eqnarray}%
to replace the Pauli operators by the fermionic operators $c_{j}$. The
Hamiltonian is transformed to a well-known Kitaev model%
\begin{eqnarray}
H_{\text{Kitaev}} &=&J\sum_{j=1}^{N-1}\left( c_{j}^{\dagger
}c_{j+1}+c_{j}^{\dag }c_{j+1}^{\dag }\right) +\text{\textrm{H.c.}}  \notag \\
&&+g\sum_{j=1}^{N}\left( 2c_{j}^{\dagger }c_{j}-1\right) .
\end{eqnarray}

To get the solution of the model, we then introduce the Majorana fermion
operators%
\begin{equation}
a_{j}=c_{j}^{\dagger }+c_{j},b_{j}=-i\left( c_{j}^{\dagger }-c_{j}\right) ,
\end{equation}%
which satisfy the commutation relations%
\begin{eqnarray}
\left\{ a_{j},a_{j^{\prime }}\right\} &=&2\delta _{j,j^{\prime }},\left\{
b_{j},b_{j^{\prime }}\right\} =2\delta _{j,j^{\prime }},  \notag \\
\left\{ a_{j},b_{j^{\prime }}\right\} &=&0.
\end{eqnarray}%
Then the Majorana representation of the original Hamiltonian is%
\begin{equation}
H_{\text{M}}=\frac{i}{2}J\sum_{j=1}^{N-1}b_{j}a_{j+1}-\frac{i}{2}%
g\sum_{j=1}^{N}a_{j}b_{j}+\text{\textrm{H.c.,}}  \label{H_SSH}
\end{equation}%
the core matrix of which is that of a $2N$-site SSH chain in a single-particle
invariant subspace. Based on the exact diagonalization result of the SSH
chain, the Hamiltonian $H_{\text{Kitaev}}$\ can be written as the diagonal
form 
\begin{equation}
H_{\text{Kitaev}}=\sum_{n=1}^{N}\varepsilon _{n}(d_{n}^{\dagger }d_{n}-\frac{%
1}{2}).
\end{equation}%
Here $d_{n}$\ is a fermionic operator, satisfying $\{d_{n},d_{n^{\prime
}}\}=0,$ and $\{d_{n},d_{n^{\prime }}^{\dag }\}=\delta _{n,n^{\prime }}$. On
the other hand, we have the relations%
\begin{equation}
\left[ d_{n},H_{\text{Kitaev}}\right] =\varepsilon _{n}d_{n},\left[
d_{n}^{\dagger },H_{\text{Kitaev}}\right] =-\varepsilon _{n}d_{n}^{\dagger },
\end{equation}%
which result in the mapping between the eigenstates of $H_{\text{Kitaev}}$.
Direct derivation shows that, for an arbitrary eigenstate $\left\vert \psi
\right\rangle $ of $H_{\text{Kitaev}}$ with eigenenergy $E$, i.e.,%
\begin{equation}
H_{\text{Kitaev}}\left\vert \psi \right\rangle =E\left\vert \psi
\right\rangle ,
\end{equation}%
state $d_{n}\left\vert \psi \right\rangle $\ $\left( d_{n}^{\dag }\left\vert
\psi \right\rangle \right) $\ is also an eigenstate of $H_{\text{Kitaev}}$\
with the eigenenergy $E-\varepsilon _{n}$ $\left( E+\varepsilon _{n}\right) $%
, i.e.,%
\begin{equation}
H_{\text{Kitaev}}\left( d_{n}\left\vert \psi \right\rangle \right) =\left(
E-\varepsilon _{n}\right) \left( d_{n}\left\vert \psi \right\rangle \right)
\end{equation}%
and%
\begin{equation}
H_{\text{Kitaev}}\left( d_{n}^{\dag }\left\vert \psi \right\rangle \right)
=\left( E+\varepsilon _{n}\right) \left( d_{n}^{\dag }\left\vert \psi
\right\rangle \right) ,
\end{equation}%
if $d_{n}\left\vert \psi \right\rangle \neq 0$ $\left( d_{n}^{\dag
}\left\vert \psi \right\rangle \neq 0\right) $.

Within the topological nontrivial region $\left\vert
g/J\right\vert <1$ ($g\neq 0$), the edge modes $d_{N}$ and $d_{N}^{\dag }$
appear with energy $\varepsilon _{N}=\pm \left\vert g/J\right\vert^{N}$. This is responsible for the fact that the ground state and the first-excited state of the Ising chain in ordered phase are quasidegenerate in finite $N$
(there is no edge mode in the trivial region $\left\vert g/J\right\vert \geqslant 1$). The
edge operator $d_{N}$ can be expressed as%
\begin{eqnarray}
d_{N} &=&\frac{1}{2}\sqrt{1-\left( \frac{g}{J}\right) ^{2}}%
\sum_{j=1}^{N}\left\{ \left[ \left( -\frac{g}{J}\right) ^{j-1}+\left( -\frac{%
g}{J}\right) ^{N-j}\right] c_{j}^{\dagger }\right.  \notag \\
&&\left. +\left[ \left( -\frac{g}{J}\right) ^{j-1}-\left( -\frac{g}{J}%
\right) ^{N-j}\right] c_{j}\right\} ,
\end{eqnarray}%
i.e., $d_{N}$ is a linear combination of particle and hole operators of
spinless fermions $c_{j}$ on the edge, and we have $[d_{N},H_{\text{Kitaev}%
}]=\varepsilon _{N}d_{N}=0$ in the large $N$ limit. Furthermore,
applying the inverse Jordan-Wigner transformation, $d_{N}$ can be expressed
as the combination of spin operators, 
\begin{eqnarray}
D_{N} &=&\frac{1}{2}\sqrt{1-\left( \frac{g}{J}\right) ^{2}}%
\sum_{j=1}^{N}\prod\limits_{l<j}\left( -\sigma _{l}^{z}\right)  \notag \\
&&\times \left[ \left( -\frac{g}{J}\right) ^{j-1}\sigma _{j}^{x}-i\left( -%
\frac{g}{J}\right) ^{N-j}\sigma _{j}^{y}\right] ,
\end{eqnarray}%
In fact, $d_{N}$\ and $D_{N}$\ are identical, but only in different
representations. Thus, from $[d_{N},H_{\text{Kitaev}}]=0$, we have 
\begin{equation}
\lbrack D_{N},H]=[D_{N}^{\dag },H]=0,  \label{commu1}
\end{equation}%
which leads to the degeneracy of the eigenstates. Furthermore, from the
canonical commutation relations $\{d_{N},d_{N}^{\dag }\}=1$ and $%
\{d_{N},d_{N}\}=0$, we have 
\begin{equation}
\{D_{N},D_{N}^{\dag }\}=1,(D_{N})^{2}=(D_{N}^{\dag })^{2}=0.  \label{commu2}
\end{equation}

For the Ising chain with slight disordered deviations on the uniform $J$ and 
$g$, the operator $D_{N}$ still exists and can be obtained by
solving the Schr\"{o}dinger equation for the corresponding SSH chain with
random hopping in the single-particle invariant subspace \cite{asboth2016short}. We have
the following solution:
\begin{equation}
D_{N}=\frac{1}{2}\sum_{j=1}^{N}\prod\limits_{l<j}\left( -\sigma
_{l}^{z}\right) \left( h_{j}^{+}\sigma _{j}^{x}-ih_{j}^{-}\sigma
_{j}^{y}\right) ,
\end{equation}
where%
\begin{eqnarray}
h_{j}^{+} &=&h_{1}^{+}\prod\limits_{m=1}^{j-1}\left( -\frac{g_{m}}{J_{m}}%
\right) ,  \notag \\
h_{j}^{-} &=&h_{N}^{-}\left( -\frac{g_{N}}{J_{j}}\right)
\prod\limits_{m=j+1}^{N-1}\left( -\frac{g_{m}}{J_{m}}\right) ,
\end{eqnarray}%
and $h_{1}^{+}$ ($h_{N}^{-}$) is determined by the normalization condition $%
\sum_{j=1}^{N}\left\vert h_{j}^{\pm }\right\vert ^{2}=1.$ The solution of $%
D_{N}$ is robust against disordered perturbation and the corresponding
energies that the edge modes are still exponentially small in $N$ under the
condition of the average value of $J_{m}$ is stronger than the average value
of $g_{m}$ \cite{asboth2016short}. Then it can be checked that the commutation
relations in Eqs. (\ref{commu1}) and (\ref{commu2}) still hold for the
operator $D_{N}$ with disordered perturbation in the large-$N$ limit.

\end{document}